# DIAGNOSTIC ANALYSIS OF SOLAR PROTON FLARES OF SEPTEMBER 2017 BY THEIR RADIO BURSTS[1]

## I.M. Chertok

*Pushkov Institute of Terrestrial Magnetism, Ionosphere and Radio Wave Propagation (IZMIRAN), Troitsk, Moscow, 108840 Russia*
*e-mail: ichertok@izmiran.ru*

**Abstract** − The powerful flares that occurred in the Sun on September 4–10, 2017 are analyzed on the basis of the technique for the quantitative diagnostics of proton flares developed in IZMIRAN in 1970–1980s. It is shown that the fluxes and energy spectra of protons with energy of tens of MeV coming to Earth qualitatively and quantitatively correspond to the intensity and frequency spectra of microwave radio bursts in the range 2.7–15.4 GHz. In particular, the flare of 4 Sept. with a peak radio flux $S \sim 2000$ sfu at the frequency $f \sim 3$ GHz (i.e., with a soft radio spectrum) was accompanied by a significant proton flux $J (> 10~\text{MeV}) \sim 100$ pfu and soft energy spectrum with the exponent $\gamma \sim 3.0$, and the powerful flare of 10 Sept. with $S \sim 21000$ sfu at $f \sim 15$ GHz (i.e., with a hard radio spectrum) led to a very intense proton event with $J (> 10~\text{MeV}) \sim 1000$ pfu with a hard energy spectrum ($\gamma \sim 1.4$), including a ground level enhancement (GLE72). This is further evidence that data on microwave radio bursts can be successfully used to diagnose proton flares, regardless of the particular source of acceleration of particles on the Sun.

## 1. INTRODUCTION

Among non-recurrent space weather disturbances, one of the most important are solar cosmic rays (SCR) or solar energetic particle events (SEPs), i.e. an increase of the proton flux in the near-Earth space with energies of tens to hundreds of MeV, and sometimes also with $E \geq 1$ GeV, recorded on terrestrial neutron monitors (see the reviews [Reames, 2013, Desai and Giacalone, 2016, Klein and Dalla, 2017]). Such enhancements can cause disruptions in operation of the electronic spacecraft equipment, lead to an increase of the radiation level at the interplanetary and near-Earth orbits, pose a danger to astronauts, as well as to crews and passengers of aircraft flying along the circumpolar routes, to lead to short-wave radio frequency disturbances at high latitudes, and so on.

Three sources are responsible for the acceleration of SEPs: a primary impulsive energy release (impulsive flare component); a prolonged post-eruptive (PE) energy release, initiated by coronal mass ejections (CMEs); a shock wave arising in the process of eruption on the front of sufficiently fast CMEs (see [Chertok, 1995; Aschwanden, 2006]). In this case, a CME itself at the meter wavelengths is manifested in the form of a type IV radio continuum, and a type II burst serves as the shock wave indicator. The PE energy release occurs at the final stage of the eruptive event, when a magnetic field in a vast region of the corona, strongly disturbed by a CME, relaxes to a new quasi-equilibrium state by means of the magnetic reconnection in an extended quasi-vertical current sheet. Depending on the CME scale and the characteristics of the coronal magnetic field, the PE phase is accompanied by considerable plasma heating, prolonged particle acceleration (sometimes up to very high energies), long-duration bursts in various ranges from gamma-rays to meter radio emission, two-ribbon structures in the chromosphere, large-scale coronal loops, giant arcades, etc. Currently, it is believed that effective and sustained particle acceleration can also occur in collapsing loops, including them at the PE phase [Somov and Kosugi, 1997].

The forecasting of SEPs is reduced to the prediction of large eruptive flares according to the characteristics of the evolving magnetic field of large sunspot groups. In addition, in a number of prognostic centers, the so-called diagnostics of proton flares is carried out. It is based on the fact that the flare electromagnetic radiation comes to the Earth for tens of minutes – hours earlier than, for example, protons with $E > 10$ MeV. This allows by the observed soft X-ray emission or radio bursts from an already occurring flare to determine in advance whether it will be accompanied by a noticeable SEP and to estimate its expected parameters [Belov et al., 2005; Kahler et al., 2007; Nûñez M., 2011; Anastasiadis et al., 2017; Zucca et al., 2017]. The original technique for the proton flare diagnostics by the accompanying radio bursts, allowing to estimate the intensity of the proton flux, its time parameters and the energy spectrum, taking into account the heliolongitude of the flare and the escape conditions of particles into the interplanetary space, was developed in IZMIRAN in 1970–1980s [Akinyan et al. , 1977, 1978, 1980a,b, 1981].

Usually SEPs occur in series due to passage through the solar disk of a large evolving active region (AR). Most SEPs are observed near the maximum of 11-year cycles, but may occur also at the declinining phase of activity. In the current 24th cycle, when approaching to the minimum, in connection with development of a large AR, on 4–10 Sept. 2017, there was an unexpectedly intense burst of activity, which included a large number of powerful flares up to an X-ray flare of the X9.3 importance and at least three significant SEPs. The present work deals with the diagnostic analysis of these SEPs based on the IZMIRAN technique mentioned above.

## 2. BASICS OF THE TECHNIQUE

The technique was based on the following proposition: the intensity parameters of microwave bursts at frequencies $f \sim 3–15$ GHz, although the latter are generated by electrons propagating to the photosphere, reflect the number of accelerated particles, including tens of MeV protons coming to the Earth. When estimating the expected intensity of the proton flux near the Earth, the initial parameters were both the maximum flux density of microwave bursts at a number of fixed frequencies and their integral parameters: total fluence and fluence of the rising phase. Thus, to a certain extent, not only the impulsive, but also the PE flare components were taken into account.

First, proton events related to flares that occurred in the so-called optimal longitude interval (OLI) were considered, within which the parameters of the proton fluxes and the character of their connection with the radio bursts are almost independent on the heliolongitude. For proton events, OLI is localized within the heliolongitudes W20–80, and its existence is due to transversal diffusion and rapid azimuthal transport of particles in the corona by means of propagation along the force lines connecting the remote active regions and areas of the Sun. Now it seems that OLI can be identified with the localization region of large-scale magnetic structures of the global solar magnetosphere that are involved in the CME eruption, as well as with the extent of the shock wave formed in front of the CME. For events in OLI, empirical relationships between the parameters of microwave bursts at several frequencies and the intensity of proton fluxes at the Earth $J$ with energy $E > 10, 30, 60$ MeV, which were called intensity functions, were established.

The introduction of the attenuation value turned out to be very fruitful, which for each event was calculated as the ratio of the proton flux observed near the Earth to one estimated by the intensity function. This made it possible to proceed to flares of arbitrary heliolongitude and, in addition, to take into account the escape conditions of particles into the interplanetary space, manifested in the meter component of radio emission (type II, IV bursts). It was assumed that the intense and weak meter component correspond to favorable and unfavorable escape conditions of particles from the flare region. Then, the distribution of events with different characteristics of meter emission on the "attenuation value of $\varphi$ – heliolongitude $\Theta$" diagrams was studied. It was established that in OLI, the events with an intense meter component (favorable escape conditions) are accompanied by more intense proton fluxes than events with a weak meter component (unfavorable escape conditions). The same effect leads to that for flares outside OLI, i.e. on the eastern half of the disk, proton fluxes in events with an intense meter component experience a stronger heliolongitude weakening than in events with a weak meter component. Analysis of the "$\varphi - \Theta$" diagrams made it possible to determine quantitatively the dependence of the weakening of the proton flux on the flare heliolongitude. Now it seems that the intensity of the meter component also reflects the power of the erupting CME and, therefore, to some extent takes into account the possible contribution of the third particle acceleration source, namely, the acceleration in the CME-driven shock. Heliongitude attenuation increases with the energy of the particles, which leads to a steepening (softening) of the energy spectrum of the protons from the eastern flares.

Moreover, in [Chertok, 1982; Chertok, 1990] (see also [Chertok et al., 2009]) it was shown that there is a direct statistical relationship between the frequency spectrum of microwave bursts and the exponent of the energy spectrum of the proton fluxes observed near the Earth with the tens of MeV energy. In particular, in OLI, flares with a soft radio spectrum (frequency of the spectral maximum $f_m \leq 5$ GHz, ratio of the peak burst intensities at 9 and 15 GHz $S_9 / S_{15} \geq 1$) lead to proton fluxes near the Earth with a soft (steep) energy spectrum, and flares with a hard radio spectrum ($f_m \geq 15$ GHz, $S_9 / S_{15} < 1$) – to proton fluxes with a hard (non-steep) energy spectrum. By the same scheme as the heliolongitude attenuation functions of the proton flux, a curve quantitatively characterizing the steepening of the proton energy spectrum for eastern events was determined.

Later, the technique was also used as a tool for investigation of SEPs. In particular, among the western events, the surplus and delayed proton fluxes were detected, in which the observed proton intensity with $E > 10$ MeV near the Earth was significantly higher than the calculated one, and the time interval between the maxima of the centimeter burst and the proton flux exceeded 10 h [Bazilevskaya et al., 1990 ]. The analysis showed that such proton fluxes are observed in events with a soft frequency spectrum of microwave bursts ($f_m \leq 5$ GHz). From this it was concluded that the delayed and surplus proton fluxes are due to the predominance of the prolonged PE acceleration, since namely the PE component of the microwave bursts is characterized by a soft radio spectrum and a considerable duration. Practically, the diagnosis of flares begins with the selection of proton events on the basis of a criterion formulated in the technique, which provided the existence of a sufficiently intense and prolonged (non-impulsive) microwave burst, as well as a substantial meter component (type II, IV bursts). In modern language, this corresponds to events with a strong flare of long duration, large CME and developed PE energy release.

## 3. FLARES AND PROTON EVENTS

A significant and very sharp outburst of solar activity in early Sept. 2017 was associated with the rapid development on the visible disk of a large sunspot group AR 12673 [Yang et al., 2017]. Fig. 1 illustrates the time histories of the soft X-ray emission in the ranges 1–8 and 0.5–4 Å and the proton flux with energies $E > 10$, 50 and 100 MeV recorded by the GOES 13 and 15 satellites during the period of 4–11 Sept. (ftp://ftp.swpc.noaa.gov/pub/warehouse/2017/2017_plots/). It can be seen that during this time, 4 X-class flares took place, including the most powerful one in the 24th cycle the X9.3 importance flare of 6 Sept., 27 M-class flares and numerous flares of lower importance. For our analysis, it is essential that all these flares occurred on the western half of the disk, in which the parameters of the proton fluxes are weakly dependent on the flare heliolongitude. According to the satellite SOHO / LASCO coronagraph (https://cdaw.gsfc.nasa.gov/CME_list/halo/halo.html), at least 3 flares ( M5.5-class on 4 Sept., X9.3-class on 6 Sept., and X8.2-class on 10 Sept. ) were accompanied by large halo CMEs. The same three powerful eruptive flares, marked by vertical arrows in Fig. 1a, were the sources of significant increases in the proton flux near the Earth, i.e. there were SEPs.

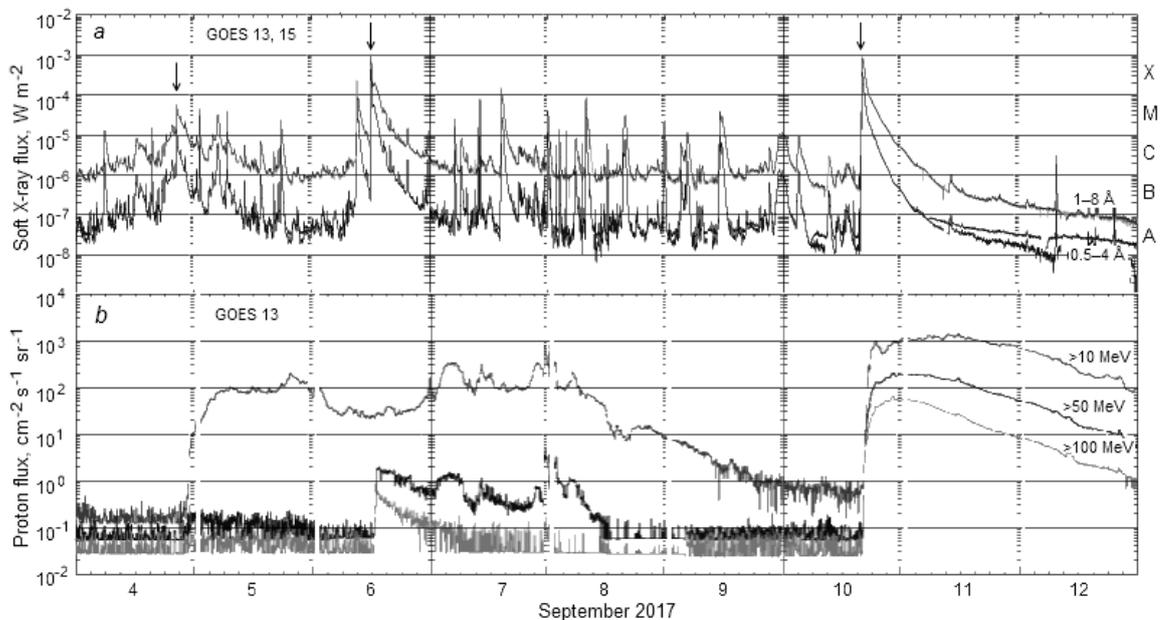

Fig. 1. Time profiles of soft X-ray emission (a) and proton fluxes (b) from the GOES satellites data for September 4-12, 2017. Vertical arrows indicate three analyzed SEPs.

As Fig. 1b shows, in the SEP of 4 Sept., the proton flux with energy $E > 10$ MeV (which we shall denote below as $J_{10}$, and at an arbitrary energy as $J_E$), enhanced to ~ 100 pfu (1 pfu = 1 $cm^{-2}s^{-1}sr^{-1}$). At the same time, an increase of the proton flux with $E > 50$ and 100 MeV was at all insignificant. It means that in this SEP, the energy spectrum of the protons was very soft (steeply falling). The index of the integral power-law energy spectrum $J_E \propto E^{-\gamma}$, which is calculated from the ratio of fluxes with $E > 10$ and 100 MeV as $\gamma = \log(J_{10}/J_{100})$, turns out to be $\gamma \sim 3.0$.

In the following days, the proton flux at $E > 10$ MeV remained elevated, probably because of the slow escape of particles from the interplanetary magnetic cloud approaching the Earth, identified with the halo CME of 4 Sept., or because of the long acceleration in the shock wave in the front of this CME. In SEP of 6 Sept. occurred on this disturbed background, a distinct increase in the proton flux is seen only at $E > 50$ and 100 MeV, where $J_{50} \sim 2$ pfu and $J_{100} \sim 0.6$ pfu. From this it follows that this event, in contrast to SEP of 4 Sept., was characterized by a rather hard (smoothly falling) spectrum of protons. Estimates by the proton fluxes at these two energies give the spectral exponent $\gamma \sim 1.7$. With such a spectrum, the flux in the $E > 10$ MeV channel should be of the order of $J_{10} \sim 32$ pfu, which is close to the protrusion observed on the proton time profile on 6 Sept., in the region of 15 UT.

The most significant SEP event took place on 10 Sept. after a long-duration near-limb flare of the X8.2 importance. The characteristics and parameters of this SEP are typical for powerful western proton flares. The proton flux of the initial fast component in this case reached $J_{10} \sim 1000$ pfu, and after a small subsequent increase it slowly subsided for several days. The flux of the fast component in the $E > 50$ and 100 MeV channels was also significant $J_{50} \sim 130$ pfu and $J_{100} \sim 40$ pfu. This indicates that its energy spectrum was somewhat harder than in the previous event, and had the exponent $\gamma \sim 1.4$. For most SEPs at energies of hundreds of MeV, the proton spectrum becomes steeper, but at $J_{10} \sim 1000$ pfu, it was enough for the flux with $E > 1$ GeV to be sufficient to register a moderate increase in terrestrial neutron monitors, i.e. such a rare event as Ground Level Enhancement (GLE) (see https://gle.oulu.fi/). This event was named as GLE72 and was only the second or third for the entire 24th cycle. Its amplitude, for example, at the IZMIRAN Moscow station was ~ 4% (http://cosrays.izmiran.ru).

## 4. RADIO BURSTS AND ANALYSIS RESULTS

In the analysis of radio bursts from the flares of this period and their use for the diagnostics of SEPs, we will proceed from data of the USAF Radio Solar Telescope Network (RSTN) providing round-the-clock observations of meter dynamic spectra and radio fluxes at several fixed frequencies in the range from 245 MHz to 15.4 GHz (ftp://ftp.sec.noaa.gov/pub/warehouse/2017/2017_events/). These tabular data allow the IZMIRAN diagnostics to be applied in a somewhat simplified form without resorting to time profiles of radio bursts, and using the maximum burst density at the frequencies 2.7 GHz ($S_3$), 5 GHz ($S_5$), 8.8 GHz ($S_9$) and 15.4 GHz ($S_{15}$).

The first stage of diagnostics consists in considering flares on the basis of the so-called protonic criterion. It was established in the IZMIRAN technique (see [Akinyan et al., 1980a]) that the flare may be a source of SEP with a proton flux near the Earth of J10 ≥ 5-10 pfu if the maximum intensity of microwave radio bursts associated with it at least at one of the frequencies in range 2.7–15.4 GHz exceeds 500 sfu (1 sfu = $10^{-22}$ W m$^{-2}$ Hz$^{-1}$). Most of the flares of the period under consideration, including those that were quite powerful, with the M-class, did not satisfy this condition and, therefore, should not lead to significant SEPs. A number of flares with the radio flux $S_{3-15}$ > 500 sfu did not correspond to another important point of the criterion: the requirement of a non-impulsive (sufficiently long) character of radio bursts. As examples of such impulsive events, one can indicate flares of M2.4-class at 05:02[2], M7.3-class at 10:15 and X1.3-class at 14:35 on 7 Sept., M8.1-class at 07:45 on 8 Sept. In these cases, the duration of microwave radio bursts was a few minutes only. The first powerful flare of X2.2-class on 6 Sept. at 09:09 is also not considered as a source of SEP due to the weakness or lack of a meter radio component. This is evidenced both by the tabular data of RSTN and the dynamic spectrum in the range of 25–270 MHz registered in IZMIRAN (see http://www.izmiran.ru/stp/lars/s_archiv.htm). This is also indicated by the fact that, according to the SOHO / LASCO coronagraph, this flare was not accompanied by a noticeable CME. In general, the conducted analysis allows us to conclude that, in accordance with observations (see Fig. 1 and Table 1), only three flares of M5.5-class at 20:48 on 4 Sept., of X9.3-class at 11:56 on 6 Sept., and of X8.2-class at15:58 on 10 Sept. completely satisfy the protonic criterion and were the sources of the three discrete SEPs described in the previous section. The increased irregular background of the proton flux on 7 and 8 Sept. seems to be due to the strong disturbance of the heliosphere and the terrestrial magnetosphere caused by propagation of a shock wave and a large CME from the X9.3-class flare of 6 Sept., the transfer of fast particles in these structures and their additional acceleration.

Fig. 2 shows the frequency spectra of three indicated flares, constructed from the maximum radio fluxes. One can immediately see that the flare of 10 Sept. possessed the most powerful radio burst at $f_m$ ~ 15 GHz ($S_{15}$ ~ 21000 sfu) and, therefore, the most rigid frequency spectrum, while the flare of 4 Sep. was accompanied by a moderate radio-burst with a maximum flux in the range of our interest at $f_m$ ~ 3 GHz ($S_3$ ~ 2000 sfu) and the softest frequency spectrum. The flare of 6 Sept. had radio bursts with intermediate characteristics: a maximum flux of $f_m$ ~ 15 GHz ($S_{15}$ ~ 8100 sfu) and a moderately hard frequency spectrum.

Estimates of the quantitative parameters of SEPs (the possible intensity scale of the proton increase in the > 10 MeV channel, $J_{10}$, and the exponent of the power energy spectrum $\gamma$) were carried out using the empirical relations obtained in development of the IZMIRAN technique and slightly adjusted later. The observed characteristics of the meter radio emission, including type II and IV bursts, allow us for all three flares to consider the conditions of the particle escape from the acceleration region to be moderate, in the terminology adopted in the IZMIRAN technique. In this case, the following empirical relationships can be used to estimate the maximum proton flux $J_{10}$ (in pfu) by the radio parameters $S_3$, $S_9$ and $S_{15}$ (in sfu) (see Akinyan et al., 1977, 1978, 1980a, b, 1981):

$$\lg(J_{10}) = 3.8\,(\lg S_3 - 3)^{1.25} + 1.6,\quad S_3 > 1000 \text{ sfu};$$

$$\lg(J_{10}) = 0.55\,(\lg S_9 - 1)^{1.14},\quad S_9 \leq 3000 \text{ sfu};$$

$$\lg(J_{10}) = 2.12\,(\lg S_9 - 3.5)^{1.729} + 1.56,\quad S_9 > 3000 \text{ sfu};$$

$$\lg(J_{10}) = 2.24\,(\lg S_{15} - 3.3)^{1.484} + 1.56,\quad S_{15} > 2000 \text{ sfu}.$$

---

[2] The article uses Universal Time (UT) everywhere.

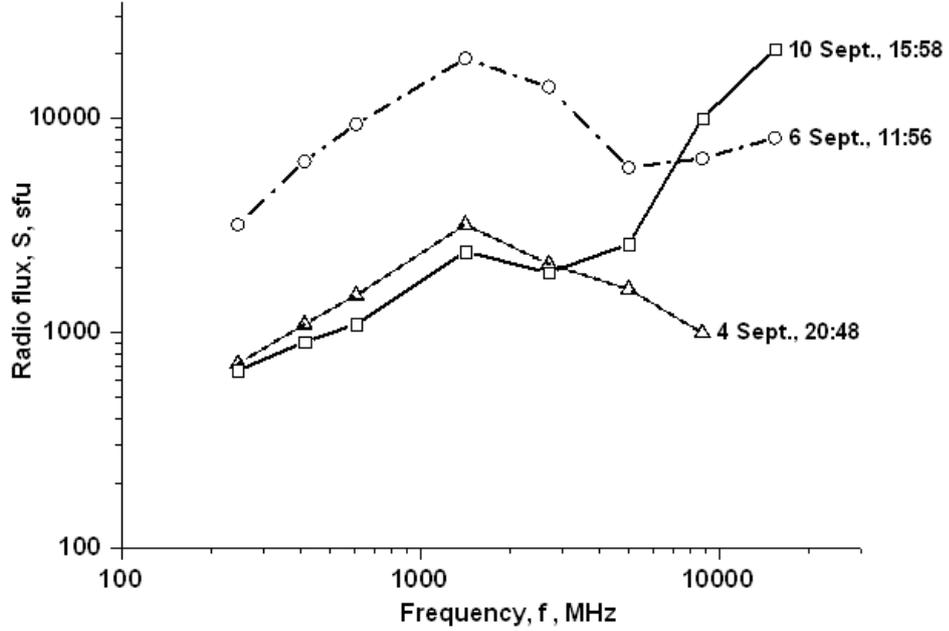

Fig. 2. Frequency spectra of radio bursts associated with the three proton events under consideration.

Estimates of the exponent of the power energy spectrum of protons $\gamma$ from the spectral maximum frequency of the radio bursts, $f_m$, and the ratio of the peak fluxes at $f \sim 9$ and 15 GHz, i.e. $S_9 / S_{15}$, are produced by expressions (see Chertok, 1982, 1990; Chertok et al., 2009):

$$\gamma = 0.91\,S_9/S_{15} + 0.432\,;$$

$$\gamma = 2.45 - 0.05\,f_m\,.$$

The results of the estimates and the corresponding observational parameters of the proton fluxes are presented in Table 1. For the flare of 4 Sept., estimates of the proton flux $J_{10}$ were carried out only by $S_3$ and $S_9$, and estimates of the exponent of the proton spectrum $\gamma$ - only by the spectral maximum frequency of the microwave burst $f_m$, since there are no data on the radio flux at $f \sim 15$ GHz for this event. For the flare of 6 Sept. the estimates of $J_{10}$ were performed over a monotonically increasing section of the frequency spectrum (see Fig. 2), i.e., by the radio parameters $S_9$ and $S_{15}$.

In general, estimates by radio data reflect quite well the intensity scale and energy spectrum of the three SEPs. A distinctive feature of the 4 Sept. flare is a soft radio spectrum with a maximum flux in the range of our interest $S_3 \sim 2000$ sfu at $f_m \sim 3$ GHz (actually the frequency spectrum reaches a peak even in the decimeter range). This determines the main estimated and observed characteristics of the proton flux: a soft energy spectrum with an exponent of $\gamma \sim 2.3$ (estimate) and $\gamma \sim 3.0$ (observations). The radio data also give a correct estimate of the SEP scale with $J_{10}$ in the range of 40–250 pfu by the observed intensity $J_{10} \sim 100$ pfu. Typical for events with a soft radio spectrum, the time profile of this SEP is characterized by a slow increase in the proton flux and a large (~ 10 h) delay of the $J_{10}$ intensity maximum relative to the flare (see [Bazilevskaya et al., 1990]). All this gives grounds to assume that in this flare the PE energy release and particle acceleration were predominant. The rise of the >10 MeV proton flux in the afternoon of 5 Sept. up to the level of ~ 200 pfu is most likely related to the contribution of CME and shock wave approaching to the Earth.

**Table 1.** Observational characteristics of the three SEPs and estimated results of estimates of the proton flux with energy > 10 MeV ($J_{10}$) by the maximum radio fluxes at frequencies ~ 3, 9, 15 GHz ($S_3$, $S_9$ and $S_{15}$), as well as the exponent of their power energy spectrum $\gamma$ by the spectral maximum frequency $f_m$ and by the ratio $S_9 / S_{15}$. The averaged values of the parameters $J_{10}$ and $\gamma$ are indicated in boldface. The peak time of microwave bursts is indicated as the flare time.

| Flares<br>Parameters | 4 Sept., 20:48<br>M5.5, S08W13 | 6 Sept., 11:56<br>X9.3, S09W38 | 10 Sept., 15:58<br>X8.2, S09W85 |
|---|---|---|---|
| $S_3$, sfu / $J_{10}$, pfu | 2000 / 270 | – | 1910 / 230 |
| $S_9$, sfu / $J_{10}$, pfu | 1000 / 40 | 6500 / 70 | 10000 / 155 |
| $S_{15}$, sfu / $J_{10}$, pfu | – | 8100 / 400 | 21000 / 7400 |
| Estimated $<J_{10}>$ | **155** | **235** | **2600** |
| Observed $J_{10}$ | **100** | **(~40)** | **1000** |
| $f_m$, GHz / $\gamma$ | 3 / 2.3 | 15 / 1.7 | 15 / 1.7 |
| $S_9/S_{15}$, $\gamma$ | – | 0.8 / 1.5 | 0.48 / 1.2 |
| Estimated $<\gamma>$ | **2.3** | **1.6** | **1.45** |
| Observed $\gamma$ | **3.0** | **1.5** | **1.4** |

Judging by the frequency radio spectrum (Fig. 2), the 6 Sept. flare was a combination of the primary flare energy release and the PE phase. The growing flux in the high-frequency part of the microwave range at $f$ ~ 5–15 GHz and a significant decimeter component with a maximum at $f$ ~ 1.4 GHz correspond to them. Estimates of the proton spectrum exponent by two microwave radio parameters indicate a rather hard proton spectrum ($\gamma$ ~ 1.6), consistent with the observed energy spectrum ($\gamma$ ~ 1.5) in the $E$> 50 and 100 MeV channels. However, the estimated > 10 MeV proton flux of $J_{10}$ ~ 235 pfu turned out to be much larger than the measured flux of $J_{10} \leq 40$ pfu (Fig. 1b). This discrepancy may be due to the disturbed state of the heliosphere, caused by the interplanetary CME from the 4 Sept. flare, which probably prevented the propagation of low-energy solar protons to the Earth. The arrival to Earth of an interplanetary magnetic cloud (ICME) with a shock wave from the 6 Sept. flare, led, on the contrary, to an increase in the proton flux at the morning and evening hours on 7 Sept.

The frequency spectrum of the 10 Sept. flare is characterized by a very sharp increase of the radio flux in the range from 3 to 15 GHz and very significant intensity $S_{15}$ ~ 21000 sfu. This leads to a hard estimated proton spectrum, consistent with the measured one of $\gamma$ ~ 1.4, and to a large scatter of estimates of the proton flux by the peak radio fluxes at different frequencies from $J_{10}$ ~ 155 pfu by $S_9$ to $J_{10}$ ~ 7400 pfu by $S_{15}$. Nevertheless, one can consider that the averaged estimated proton flux $J_{10}$ ~ 2600 pfu indicates a fairly correct the SEP scale. The latter for the fast component was $J_{10}$ ~ 1000 pfu, and for the delayed component it increased by a factor of ~ 1.4. It is significant that the estimates of the intensity and spectrum of the proton fluxes, in particular the significant $J_{10}$ value, obtained by radio data, corresponds to the observed ground enhancement (GLE) in this case.

## 5. CONCLUSIONS

The analysis makes it possible to conclude that the application of the technique of the quantitative radio diagnostics of proton flares, developed in IZMIRAN in 1970-1980s, to a series of powerful flares of Sept. 2017 leads, on the whole, to positive results. First of all, on the basis of the proton flare criterion provided for in the technique, which takes into account the data on the intensity, duration, frequency spectrum of microwave bursts, and the parameters of the meter component of radio emission, flares of different character were correctly identified, which were the sources of three significant SPEs. Estimates of the possible proton flux intensity in the $E > 10$ MeV channel, carried out in a simplified version by the peak radio flux at frequencies $f \sim 3, 9$ and 15 GHz, made it possible to determine fairly accurately the scale of all three SEPs. By type of the frequency spectrum of microwave radio bursts, it was successfully diagnosed both the flare with the predominant PE component (4 Sept.) and two flares in which the main contribution to the proton acceleration was apparently given by the primary flare energy release (6 and 10 Sept.).

Particularly important is the possibility demonstrated above to estimate correctly the exponent of the proton energy spectrum in the range of tens of MeV by the radio spectrum. It was again shown that a PE flare with a soft frequency spectrum of microwave bursts causes SEP with a soft energy proton spectrum, and flares with a hard radio spectrum, on the contrary, are accompanied by proton fluxes with a hard energy spectrum. Obviously, this correspondence between radio and energy spectra does not fit into the so-called "large flare syndrome" [Kahler, 1982], according to which the proton flux intensity near the Earth shows a noticeable positive correlation with any parameters reflecting the flare energy, including those unrelated directly physically to the particle acceleration.

The obtained results indicate that the flare particle acceleration, which includes the primary acceleration and acceleration in the PE current sheet, formed in the corona after CME, makes a significant contribution to the proton fluxes observed near the Earth. The carried out research once again demonstrates that, regardless of specific sources of acceleration, the data on radio bursts can be successfully used both for analysis and for real diagnostics of proton flares, i.e. for preliminary estimates of the intensity and energy spectrum of proton fluxes. It appears to be rather promising the flare diagnostics by radio bursts at higher frequencies $\sim 35$ GHz and the use wherein fluences of microwave bursts as a primary initial parameter [Grechnev et al, 2013, 2017].

**Acknowledgments** The author thanks the NOAA/SWPC GOES and USAF RSTN teams for the open data used in the study. This research was partially supported by the Russian Foundation of Basic Research under grant 17-02-00308.